\documentclass[pre,twocolumn,eqsecnum,superscriptaddress]{revtex4-1}
\usepackage{epsfig, graphicx, color}
\usepackage{mathrsfs}
\usepackage{amssymb,amsmath}
\newcommand{\vt}[1]{{\mathbf #1}}

\newcommand{\oH}{\overline{H}}

\newcommand{\beq}{\begin{equation}}
\newcommand{\eeq}{\end{equation}}

\begin{document}

\title{Fractional Brownian motion and generalized Langevin equation motion in
confined geometries}

\author{Jae-Hyung Jeon}
\email[]{jae-hyung.jeon@ph.tum.de}
\author{Ralf Metzler}
\email[]{metz@ph.tum.de}
\affiliation{Department of Physics, Technical University of Munich,
James-Franck Stra{\ss}e, 85747 Garching, Germany}

\begin{abstract}
Motivated by subdiffusive motion of bio-molecules observed in
living cells we study the stochastic properties of a non-Brownian
particle whose motion is governed by either fractional Brownian
motion or the fractional Langevin equation and restricted to a
finite domain. We investigate by analytic calculations and
simulations how time-averaged observables (e.g., the time averaged
mean squared displacement and displacement correlation) are
affected by spatial confinement and dimensionality. In particular
we study the degree of weak ergodicity breaking and scatter
between different single trajectories for this confined motion in
the subdiffusive domain. The general trend is that deviations from
ergodicity are decreased with decreasing size of the movement
volume, and with increasing dimensionality. We define the
displacement correlation function and find that this quantity
shows distinct features for fractional Brownian motion, fractional
Langevin equation, and continuous time subdiffusion, such that it
appears an efficient measure to distinguish these different
processes based on single particle trajectory data.
\end{abstract}

\maketitle

\section{Introduction}

Anomalous diffusion denominates deviations from the regular linear growth
of the mean squared displacement $\langle\mathbf{r}^2(t)\rangle\simeq K_1
t$ as a function of time $t$, where the proportionality factor $K_1$ is
the diffusion constant of dimension $\mathrm{cm}^2/\mathrm{sec}$. Often
these deviations are of power-law form, and in this case the mean squared
displacement in $d$ dimensions
\begin{equation}
\label{msd}
\langle\mathbf{r}^2(t)\rangle=\frac{2dK_{\alpha}}{\Gamma(1+\alpha)}t^{\alpha}
\end{equation}
describes subdiffusion when the anomalous diffusion exponent is in the range
$0<\alpha<1$, and superdiffusion for $\alpha>1$ \cite{report}. The generalized
diffusion constant $K_{\alpha}$ has dimension $\mathrm{cm}^2/\mathrm{sec}^{
\alpha}$. We here concentrate on subdiffusion phenomena. Power-law mean
squared displacements of the form (\ref{msd}) with $0<\alpha<1$ have been
observed in a multitude of systems such as amorphous semiconductors
\cite{scher}, subsurface tracer dispersion
\cite{grl}, or in financial market dynamics \cite{mainardi}. Subdiffusion
is quite abundant in small systems. These include the motion of small probe
beads in actin networks \cite{weitz}, local dynamics in polymer melts
\cite{kimmich}, or the motion of particles in colloidal glasses \cite{weeks}.
\emph{In vivo}, crowding-induced subdiffusion has been reported for RNA motion in \emph{E.coli}
cells \cite{golding}, the diffusion of lipid granules embedded in the
cytoplasm \cite{elbaum,lene}, the propagation of virus shells in cells
\cite{seisenhuber}, the motion of telomeres in mammalian cells \cite{garini},
as well as the diffusion of membrane proteins and of dextrane probes in
HeLa cells \cite{weiss}.

While the mean squared displacement Eq.~(\ref{msd}) is commonly used to classify
a process as subdiffusive, it does not provide any information on the physical
mechanism underlying this subdiffusion. In fact there are several pathways
along which this subdiffusion may emerge. The most common are:

(i) The continuous time random walk (CTRW) \cite{scher} in which the step length has
a finite variance $\langle\delta\mathbf{r}^2\rangle$ of jump lengths, but the
waiting time $\tau$ elapsing between successive jumps is distributed as a power
law $\psi(\tau)\simeq\tau_0^{\alpha}/\tau^{1+\alpha}$, with $0<\alpha<1$. The
diverging characteristic waiting time gives rise to subdiffusion of the form
(\ref{msd}) \cite{scher}. The subdiffusive CTRW in the
diffusion limit is equivalent to the fractional Fokker-Planck equation, that
directly shows the long-ranged memory intrinsic to the process \cite{report}.
Waiting times of the form $\psi(\tau)$ were, for instance, observed in the
motion of tracer beads in an actin network \cite{weitz}. We note that
recently subdiffusion was also demonstrated in a coupled CTRW model \cite{vincent}.

(ii) A random walk on a fractal support meets bottlenecks and dead ends on all
scales and is subdiffusive. The resulting subdiffusion is also of
the form (\ref{msd}), and the anomalous diffusion exponent is related to the
fractal and spectral dimensions, $d_f$ and $d_s$, characteristic of the
fractal, through $\alpha=d_s/d_f$ \cite{havlin}. A typical example is the
subdiffusion on a percolation cluster near criticality that was actually
verified experimentally \cite{kimmich1}.

(iii) Fractional Brownian motion (FBM) and the fractional Langevin
equation (FLE) that will be in the focus of this work and will be
defined in Sec.~\ref{sec2}. The understanding of these types of
stochastic motions is up to date somewhat fragmentary. Thus the
first passage behavior of FBM is known analytically only in one
dimension on a semi-infinite domain \cite{molchan}; the escape
from potential wells in the framework of FLE has been studied
analytically \cite{goychuck,chaudhury} and numerically
\cite{chechkin,goychuck2}; and a priori unexpected critical
exponents have been identified for the FLE \cite{stas1}. Here we
address a fundamental question related to FBM and FLE. Namely,
what is their behavior under confinement? Two main aspects of this
question will be addressed. One is the study of the relaxation
towards stationarity in a finite box by means of the ensemble
averaged mean squared displacement. We also investigate a new
quantity used to characterize the motion, the displacement
correlation function. For these aspects we also study the
dependence on the dimensionality of the motion.

The second aspect concerns how FBM and FLE motions under confinement behave
with respect to ergodicity.
Experimentally the recording of single particle trajectories has become a
standard tool, producing time series of data that are then analyzed by
time rather than ensemble averages. For subdiffusion processes both are
not necessarily identical. In fact for CTRW subdiffusion with an ensemble
averaged mean squared displacement of the form (\ref{msd}) the time averaged
mean squared displacement
\begin{equation}
\label{tamsd}
\overline{\delta^2(\Delta,T)}=\frac{1}{T-\Delta}\int_0^{T-\Delta}\big[x(t+
\Delta)-x(t)\big]^2dt
\end{equation}
on average scales like
$\left<\overline{\delta^2(\Delta,T)}\right>\simeq\Delta/T^{1-
\alpha}$ \cite{YHe,app}. That is, the anomalous diffusion is
manifested only in the dependence on the overall measurement time
$T$ and not in the lag time $\Delta$, that defines a window swept
along the time series. Thus time and ensemble averages are indeed
different. In contrast, for normal Brownian diffusion
$\left<\overline{\delta^2(\Delta,T)}\right>\simeq\Delta$ is
independent of $T$, and time and ensemble averages become
identical, i.e., the system is ergodic. Different from CTRW
subdiffusion, systems governed by FBM or FLE are ergodic. The
ergodicity breaking parameter measured from time averaged mean
squared displacements converges algebraically to zero [ergodic
behavior] as the measurement time increases, the convergence speed
depending on the Hurst exponent $H=\alpha/2$ \cite{deng}. For the
FLE case, however, it was also shown that the ergodicity measured
from the velocity variance can be broken for a class of colored
noises \cite{bao}.

One of the open questions in the context of ergodicity breaking for FBM and
FLE in the above sense is the influence of boundary conditions on the time
averages. It was shown for CTRW subdiffusion that confinement changes the
short time scaling $\left<\overline{\delta^2(\Delta,T)}\right>\simeq\Delta/
T^{1-\alpha}$ to the long time behavior $\left<\overline{\delta^2(\Delta,T)}
\right>\simeq(\Delta/T)^{1-\alpha}$ \cite{stas,thomas}. Although we expect that
FBM and FLE processes become stationary under confinement and, for instance,
attain the same long-time mean squared displacement dictated by the size of
the confinement volume, we investigate how fast this relaxation actually is,
and how it depends on the volume and the dimensionality. To this end we study
the ergodicity breaking parameter for the system. We find that for both FBM
and FLE confinement actually decreases the value of the ergodicity breaking
parameter with respect to unbounded motion, i.e., the process becomes more
ergodic. Ergodicity is also enhanced with increasing dimensionality. We also
discuss how FBM and FLE motions can be distinguished from time series from
single particle trajectories.

The paper is organized as follows. In Sec.~\ref{sec2}, we introduce FBM and
FLE motions, and review briefly their basic statistical properties. In
Sec.~\ref{sec3}, we describe the numerical scheme for simulating FBM and FLE
in confined space. Simulations results are presented in Sec.~\ref{sec4} and
\ref{sec5}, where we discuss the effects of confinement and dimensionality on
time-averaged mean squared displacement trajectory, ergodicity, and displacement correlation. We draw our Conclusions in Sec.~\ref{sec6}.

\section{Theoretical model}
\label{sec2}

We here define FBM and FLE. These two stochastic models share many common
features, however, their physical nature is different. In the following we
will see, in particular, how the two can be distinguished on the basis of
experimental or simulations data.

\subsection{Fractional Brownian motion}

FBM was originally introduced by Kolmogorov in 1940 \cite{kolmogorov} and
further studied by Yaglom \cite{yaglom}. In a different context it was
introduced by Mandelbrot in 1965 \cite{mandelbrot1} and fully
described by Mandelbrot and van Ness in 1968 in terms of a stochastic
integral representation \cite{mandelbrot}. In the latter
reference the authors wrote that "We believe FBMs do provide useful models for
a host of natural time series". This study was motivated by Hurst's analysis
of annual river discharges \cite{hurst}, the observation that in economic time
series cycles of all orders of magnitude occur \cite{finance}, and that many
experimental studies exhibit the now famed $1/f$ noise. FBM by now is widely
used across fields. Among many others FBM has been identified as the underlying
stochastic process of the subdiffusion of large molecules in biological cells
\cite{weiss,weiss1,vincent1}. We note that FBM is
neither a semimartingale nor a Markov process, which makes it quite intricate
to study with the tools of stochastic calculus \cite{oksendal,weron}.

FBM, $x^H(t)$, is a Gaussian process with stationary increments which
satisfies the following statistical properties: the process is symmetric,
\begin{equation}
\langle x^H(t)\rangle=0,
\end{equation}
with $x^H(0)=0$; and the second moment scales like Eq.~(\ref{msd}):
\begin{equation}
\langle x^H(t)^2\rangle=2K_Ht^{2H}.
\end{equation}
For easier comparison with other literature we introduced the Hurst exponent
$H$ that is related to the anomalous diffusion exponent via $H=\alpha/2$. The
Hurst exponent may vary in the range $0<H<1$, such that FBM describes both
subdiffusion ($0<H<1/2$) and superdiffusion ($1/2<H<1$). The limits $H=1/2$
and $H=1$ correspond to Brownian and ballistic motion, respectively. Finally, the two
point correlation behaves as
\begin{equation}
\langle x^H(t_1)x^H(t_2)\rangle=K_H(t_1^{2H}+t_2^{2H}-|t_1-t_2|^{2H}).
\label{eq:covar-fbm}
\end{equation}
Here $\langle\cdot\rangle$ represents the ensemble average. It is convenient
to introduce fractional Gaussian noise (FGN), $\xi^H(t)$, from which the FBM
is generated by
\begin{equation}
x^H(t)=\int_0^tdt'\xi^H(t').
\end{equation}
FGN has the properties of zero mean
\begin{equation}
\langle \xi^H(t)\rangle=0
\end{equation}
and autocorrelation~\cite{qian2,xie}
\begin{eqnarray}
\label{xi}
\langle\xi^H(t_1)\xi^H(t_2)\rangle&=&2K_HH(2H-1)|t_1-t_2|^{2H-2}\nonumber \\
&+&4K_HH|t_1-t_2|^{2H-1}\delta(t_1-t_2),\nonumber \\
\end{eqnarray}
as can be seen by differentiation of Eq.~(\ref{eq:covar-fbm}) with
respect to $t_1$ and $t_2$. Here we see that $K_H$ plays the role
of a noise strength. For subdiffusion ($0<H<1/2$) the
autocorrelation is negative for $t_1\neq t_2$, i.e., the process
is anti-correlated or antipersistent~\cite{FGNcomment}. In
contrast, for $1/2<H<1$, the noise is positively correlated
(persistent) and the motion becomes superdiffusive. For normal
diffusion ($H=1/2)$, the noise is uncorrelated, i.e.,
$\langle\xi^H(t_1)\xi^H(t_2)\rangle=2K_H\delta(t_1-t_2)$. For
further details compare the discussions in
Ref.~\cite{mandelbrot,feder}.

We define $d$-dimensional FBM as a superposition of independent FBMs for each
Cartesian coordinate, such that
\begin{equation}
\label{eq:fbm-2d3d}
\vt{x}^H(t)=\sum_{i=1}^d \int_0^tdt'\xi_i^H(t')\hat{x}_i,
\end{equation}
where $\hat{x}_i$ is the Cartesian coordinate of the $i$th component and
$\xi_i^H$ is FGN which satisfies
\begin{equation}
\langle \xi_i^H(t)\rangle=0
\end{equation}
and
\begin{eqnarray}
\langle\xi_i^H(t_1)\xi_j^H(t_2)\rangle&=&2K_HH(2H-1)|t_1-t_2|^{2H-2}\delta_{ij}\nonumber
\\
&+&4K_HH|t_1-t_2|^{2H-1}\delta(t_1-t_2)\delta_{ij}.\nonumber \\
\end{eqnarray}
From this definition, $d$-dimensional FBM $\vt{x}^H(t)$ has the
properties of zero mean
\begin{equation}
\langle \vt{x}^H(t)\rangle=0,
\end{equation}
variance
\begin{equation}
\langle \vt{x}^H(t)^2\rangle=2dK_Ht^{2H},
\end{equation}
and autocorrelation
\begin{equation}
\langle\vt{x}^H(t_1)\cdot\vt{x}^H(t_2)\rangle=dK_H(t_1^{2H}+t_2^{2H}-|t_1
-t_2|^{2H}).
\end{equation}
Note that $|\vt{x}^H(t)|$ cannot satisfy these properties, thus it is not
an FBM.

A few remarks on this multidimensional extension of FBM are in
order. We note that, albeit intuitive due to the Gaussian nature
of FBM, this multidimensional extension is not necessarily unique.
In mathematical literature higher dimensional FBM in the above
sense was defined in Refs.~\cite{unterberger,qian}. In physics
literature an analogous extension to higher dimensions was used in
Ref.~\cite{weiss} based on the Weierstrass-Mandelbrot method. To
verify that this $d$-dimensional extension is meaningful, we
checked from our simulations of $d$-dimensional FBM that the
fractal dimension of FBM, $d_f=1/H$ for $H>1/d$ \cite{falconer}, is preserved in higher dimensions. Moreover we used an alternative method to create FBM in
$d$-dimensions, namely, to use 1D FBM to choose the length  of a
radius and then choose the space angle randomly. The results were
equivalent to the above definition to use independent FBMs for
every Cartesian coordinate. We are therefore confident that our
definition of FBM in $d$ dimensions is a proper extension of
regular 1D FBM.

\subsection{Fractional Langevin equation motion}

An alternative approach to Brownian motion is based on the Langevin equation
\cite{langevin,vankampen,coffey}
\begin{equation}
m\frac{d^2y(t)}{dt^2}=-\overline{\gamma}\frac{dy(t)}{dt}+\xi(t),
\end{equation}
where $\xi(t)$ corresponds to white Gaussian noise.

When the random noise $\xi(t)$ is non-white, the resulting motion is described
by the generalized Langevin equation (GLE)
\begin{eqnarray}
m\frac{d^2y(t)}{dt^2}=-\overline{\gamma}\int_0^t\mathscr{K}(t-t')\frac{dy}{dt'}
dt'+\xi(t),
\label{eq:GLE}
\end{eqnarray}
where $m$ is the test particle mass, $\mathscr{K}$ is the memory kernel
\cite{zwanzig,berne,kubo} which
satisfies the fluctuation-dissipation theorem $\langle\xi(t)\xi(t')\rangle=
k_BT\overline{\gamma}\mathscr{K}(t-t')$. When $\xi$ is the FGN introduced
above, $\mathscr{K}$ decays algebraically, and Eq.~(\ref{eq:GLE})
becomes the fractional Langevin equation (FLE)
\begin{eqnarray}
\nonumber
m\frac{d^2y(t)}{dt^2}&=&-\overline{\gamma}\int_0^t(t-t')^{2\overline{H}-2}\frac{dy}{dt'}
dt'+\eta\xi^{\overline{H}}(t)\\
&&\hspace*{-1.2cm}=-\overline{\gamma}\Gamma(2\overline{H}-1)\frac{d^{2-2\overline{H}}}
{dt^{2-2\oH}}
y(t)+\eta\xi^{\oH}(t).
\label{eq:FLE1}
\end{eqnarray}
Here, $\overline{\gamma}$ is a generalized friction coefficient. We also
define the coupling constant
\begin{equation}
\eta=\sqrt{\frac{k_BT\overline{\gamma}}{2K_H \oH(2\oH-1)}}
\end{equation}
imposed by the fluctuation dissipation theorem, and the Caputo fractional
derivative \cite{podlubny}
\begin{equation}
\frac{d^{2-2\oH}}{dt^{2-2\oH}}y(t)=\frac{1}{\Gamma(2\oH-1)}\int_0^tdt'(t-t')^{2\oH-2}
\frac{dy}{dt'}.
\end{equation}
Note that in Eq.~(\ref{eq:FLE1}) the memory integral diverges for
$\oH$ smaller than $1/2$, such that the Hurst exponent in the FLE
is restricted to the range $1/2<\oH<1$.


It can be shown that the relaxation dynamics governed by the FLE
(\ref{eq:FLE1}) follows the form
\begin{equation}
\langle y(t)\rangle=v_0t E_{2\oH,2}\left(-\gamma t^{2\oH}\right)
\end{equation}
for the first moment, where $v_0$ is the initial particle
velocity. The rescaled friction coefficient is
$\gamma=\overline{\gamma}\Gamma(2\oH-1)/m$. The coordinate
variance behaves as
\begin{equation}
\langle y^2(t)\rangle=2\frac{k_BT}{m}t^2E_{2\oH,3}\left(-\gamma t^{2\oH}\right)
\label{eq:msd-FLE}
\end{equation}
where $\langle v_0^2\rangle=k_BT/m$ is assumed, and we employed
the generalized Mittag-Leffler function \cite{bateman}
\begin{equation}
E_{\alpha,\beta}(z)=\sum_{n=0}^{\infty}\frac{z^n}{\Gamma(\alpha n+\beta)}
\end{equation}
whose asymptotic behavior for large $z$ is
\begin{equation}
E_{\alpha,\beta}(z)=-\sum_{n=1}^{\infty}\frac{z^{-n}}{\Gamma(\beta-\alpha n)}.
\end{equation}
Thus the mean squared displacement shows a turnover from short
time ballistic motion to long time anomalous diffusion of the
form~\cite{lutz,pottier}
\begin{equation}
\langle y^2(t)\rangle\sim\left\{\begin{array}{ll} t^2, & t\to0\\
t^{2-2\oH}, & t\to\infty\end{array}\right..\label{FLEscaling}
\end{equation}
Therefore, for persistent noise with $1/2<\oH<1$ the resulting
motion is in fact subdiffusive, i.e., the persistence of the noise
has the opposite effect than in FBM.

In analogy to our discussion of FBM in a $d$-dimensional embedding the FLE
is generalized to
\begin{eqnarray}
m\frac{d^2\vt{y}(t)}{dt^2}=-\overline{\gamma}\int_0^t\Vec{\Vec{\mathscr{K}}}
\cdot\frac{d\vt{y}}{dt'}dt'+\mbox{\boldmath${\xi}$}(t),
\end{eqnarray}
where $\vt{y}(t)=\sum_iy_i(t)\hat{x}_i$, $\mbox{\boldmath${\xi}$}(t)=\sum_{i}
\xi_i^{H}(t)\hat{x}_i$, and $\Vec{\Vec{\mathscr{K}}}$ is the memory tensor
which is in diagonal form (i.e., $\mathscr{K}_{ij}=\mathscr{K}(t-t')
\delta_{ij}$) in the absence of motional coupling between different coordinates.

\section{Simulations scheme}
\label{sec3}

We here briefly review the simulations scheme used to produce time series for
FBM and FLE motion.

\subsection{Fractional Brownian motion}

$d$-dimensional FBM is simulated via Eq.~(\ref{eq:fbm-2d3d}) by
numerical integration of $\xi_i^H(t)$. The underlying FGN was
generated by the Hosking method which is known to be an exact but
time-consuming algorithm \cite{hosking}. We checked that in the
one-dimensional case the generated FBM in free space successfully
reproduces the theoretically expected behavior, the mean squared
displacement (\ref{msd}), the fractal dimension $d_f=2-\alpha/2$ of
the resulting trajectory, and the first passage time distribution.
To simulate the confined motion, reflecting walls were considered
at locations $\pm L$ for each coordinate. For instance in the  1D
case, if $|x^H(t)|>L$ at some time $t$, the particle bounces back
to the position $x^H(t)-2|x^H(t)-\mathrm{sign}(x^H)L|$. Similar
reflecting conditions were taken into account in the
multi-dimensional case.

\subsection{Fractional Langevin equation motion}

In simulating FLE motion, we follow the numerical method presented
by Deng and Barkai~\cite{deng}. First, integrating
Eq.~(\ref{eq:FLE1}) from 0 to $t$, we obtain the Volterra integral
equation for velocity field $v(t)=dy(t)/dt$
\begin{eqnarray}
\nonumber
v(t)&=&-\frac{\overline{\gamma}}{(2\oH-1)m}\int_0^t(t-t')^{2\oH-1}v(t')dt'\\
&&+v_0+\frac{\eta}{m}x^{\oH}(t),
\end{eqnarray}
where $v(t=0)=v_0$. This stochastic integral equation can be
evaluated by the predictor-corrector algorithm presented in
Ref.~\cite{kai} with the FBM $x^{\oH}(t)$ independently obtained
by the Hosking method. We calculated $y(t)=y_0+\int_0^t v(t')dt'$
by the trapezoidal algorithm. For discrete time steps, the
equation of motion is given by
\begin{eqnarray}
\nonumber
y_{n+1}&=&y_0+\frac{dh}{2}(v_0+v_{n+1})+dh\sum_{i=1}^{n}v_i,\\
&=&\frac{dh}{2}(v_n+v_{n+1})+y_n,
\label{eq:GLE-numerical}
\end{eqnarray}
where $dh$ is the time increment. When evaluating
Eq.~(\ref{eq:GLE-numerical}), a reflecting boundary condition was
considered in the sense that $y_n\rightarrow
y_n-2|y_n-\hbox{sign}(y_n)L|$ if $|y_n|>L$.

\begin{figure}
\includegraphics[width=0.50\textwidth]{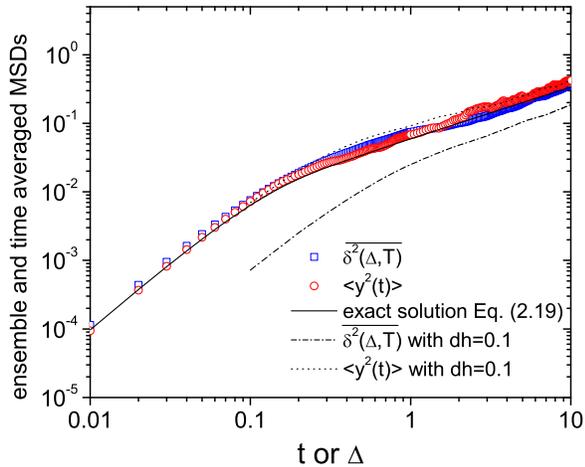}
\caption{(Color online) The mean squared displacements (MSD) for FLE motion in free space.
The ensemble averaged and time averaged MSDs, $\langle y^2(t)\rangle$ and
$\overline{\delta^2(\Delta,T)}$, obtained from simulation are compared with
the exact solution $2t^2E_{5/4,3}[-10\Gamma(5/4-1)t^{5/4}]$
given by Eq.~(\ref{eq:msd-FLE}). The ensemble averaged value was obtained from 200 simulated trajectories.
In the simulation, we chose the Hurst exponent $\oH=5/8$, time increment $dh=0.01$, particle mass $m=1$, initial velocity
$v_0=1$, initial position $y_0=0$, friction coefficient $\overline{\gamma}=10$,
and $k_BT=1$.}
\label{f:nerror}
\end{figure}
To show the reliability of our simulation, we compare the
simulation result with the well-known solution for free space
motion. In Fig. 1, we simulate the subdiffusion case with the parameters
values, $\oH=5/8$, $m=1$, $v_0=1$, $y_0=0$,
$\overline{\gamma}=10$, $k_BT=1$, and $dh=0.01$. From 200
simulated trajectories, we obtain the ensemble averaged and time
averaged mean squared displacements, $\langle y^2(t)\rangle$ and
$\overline{\delta^2(\Delta,T)}$, and compare them with the exact
solution Eq.~(\ref{eq:msd-FLE}). Note that $\langle y^2(t)\rangle$
should be identical to $\overline{\delta^2(\Delta,T)}$,
Eq.~(\ref{eq:tmsd-FLE}), with $t$ regarded as the lag time
$\Delta$ due to the ergodicity of the FLE motion in free
space~\cite{deng}. The deviation from the exact solution is
markedly reduced with decreasing time increment $dh$. With our
chosen value $dh=0.01$ the mean squared displacements obtained
from simulation appear to be in good agreement with the theory.

\section{Fractional Brownian Motion in confined space}
\label{sec4}

We now turn to the investigation of the behavior of FBM under
confinement, analyzing the mean squared displacement and potential
ergodicity breaking. We then define the displacement correlation
function, and finally study the influence of dimensionality.

\subsection{Mean squared displacement}

For FBM in free space $\langle x^H(t)^2\rangle$ can be estimated by the time
averaged mean squared displacement Eq.~(\ref{tamsd}) via the exact relation
\cite{deng}
\begin{equation}
\left<\overline{\delta^2(\Delta,T)}\right>=2K_H\Delta^{2H}.
\end{equation}
Here $\langle\cdot\rangle$ denotes the ensemble average. In contrast to
CTRW subdiffusion, in FBM this quantity is ergodic. However, as mentioned
above, the approach to ergodicity is algebraically slow, and we want to
explore here whether boundary conditions have an impact on the ergodic
behavior. Let us now analyze the behavior in a box of size $2L$.

\begin{figure}
\includegraphics[width=0.50\textwidth]{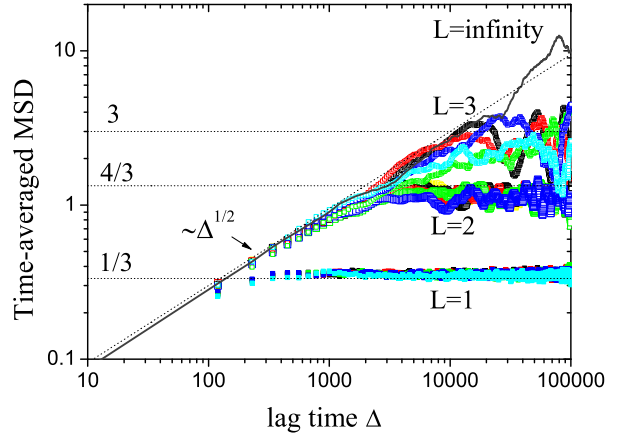}
\caption{(Color online) Time-averaged mean squared displacement (MSD) versus the lag time
$\Delta$ for given values of $L$. The drawn line has slope $1/2$,
corresponding to the expected short lag time behavior for the used value
$H=1/4$ of the Hurst exponent. For $L=1$, 2, and 3, five different
trajectories each are drawn to be able to see whether the trajectories
scatter. The simulation time is $T=2^{17}\approx1.3\cdot10^5$.}
\label{f:msd-fbm}
\end{figure}

In Fig.~\ref{f:msd-fbm} we show typical curves for the time averaged mean
squared displacement for Hurst exponent $H=1/4$ and three different interval
lengths $L$. Regardless of the size of $L$, the confined environment does not
affect the power law with exponent $2H$ for short lag times. Moreover at long
lag times we observe saturation of the curves to a value that depends on $L$.
This behavior is distinct from that of the CTRW case where $\left<\overline{
\delta^2(\Delta,T)}\right>$ shows a power law with slope $1-\alpha$ \cite{stas,thomas}.

One can estimate the saturated value as a function of $L$. For long $\Delta$
and measurement time $T$, the probability $p(x)$ to find the particle located
at $x$ is independent of $x$ due to the equilibration between the reflecting
walls, and thus $\int_{-L}^{L} x^2p(x)dx=L^2/3$. The dotted lines in
Fig.~\ref{f:msd-fbm} represent these values.

We observe that the scatter between different single trajectories becomes more
pronounced when the interval length is increased. In fact the scatter is
negligible for $L=1$ while it is quite appreciable for $L=3$, even though
the slope of all curves at finite $L$ converges to a horizontal slope, with
an amplitude close to the predicted value $L^2/3$. We also note that the
scatter depends on the total measurement time $T$. For given $L$ it tends to
be reduced as we increase $T$. This effect will be discussed quantitatively
in detail using the ergodicity breaking parameter.

\subsection{Ergodicity breaking parameter}

In contrast to CTRW subdiffusion, FBM in free space is known to be ergodic
\cite{deng}. The time averaged mean squared displacement traces displayed
in Fig.~\ref{f:msd-fbm} exhibit no extreme scatter as known from the CTRW
case. This implies that ergodicity is indeed preserved for confined FBM. We
quantify this statement more precisely in terms of the ergodicity breaking
parameter \cite{YHe}
\begin{eqnarray}
E_B(\Delta,T)=\frac{\left<\left(\overline{\delta^2(\Delta,T)}\right)^2\right>
-\left<\overline{\delta^2(\Delta,T)}\right>^2}{\left<\overline{\delta^2(\Delta
,T)}\right>^2},
\label{eq:EB}
\end{eqnarray}
where $\lim_{T\rightarrow\infty}E_B(T)=0$ is expected for ergodic
systems. For the case of free FBM, Deng and Barkai analytically
derived that $E_B$ decays to zero as
\begin{eqnarray}
E_B(\Delta,T)\sim\left\{
\begin{array}{ll}\displaystyle\frac{\Delta}{T} & \mbox{for}~ 0<H<\frac{3}{4},\\[0.4cm]
\displaystyle\frac{\Delta}{T}\log T & \mbox{for}~ H=\frac{3}{4}, \\[0.4cm]
\displaystyle  \left(\frac{\Delta}{T}\right)^{4-4H} & \mbox{for}~\frac{3}{4}<H<1,
\end{array}\right.
\end{eqnarray}
for long measurement time $T$~\cite{deng}.

\begin{figure}
\includegraphics[width=0.50\textwidth]{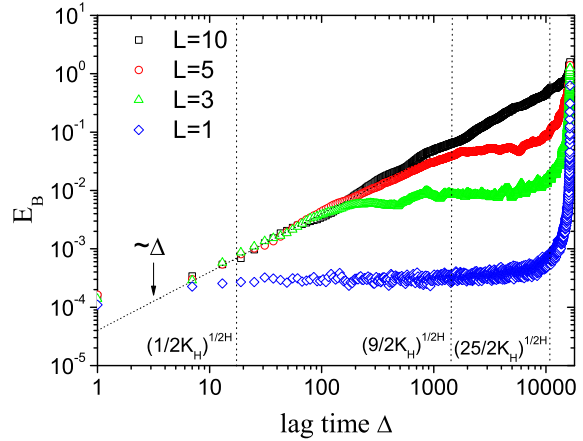}
\caption{(Color online) Ergodicity breaking parameter $E_B$ versus lag time $\Delta$ for
$L=1$, 3, 5, and 10 (from bottom to top) with Hurst exponent $H=1/4$. The
overall measurement time is $T=2^{14}\approx1.6\cdot10^4$. The dotted line with slope 1
represents the theoretical expectation $E_B\simeq\Delta$ in free space. For
each $L$ the curve was obtained from 200 single trajectories. The vertical lines show the
crossover lag time $\Delta_{cr}=(L^2/2K_H)^{1/2H}$ for $L=1,~3$, and 5. }
\label{f:EB-fbm-lag}
\end{figure}

We numerically investigate the boundary effects on the ergodicity
breaking parameter. First, in Fig.~\ref{f:EB-fbm-lag} we evaluate
$E_B$ as function of the lag time $\Delta$ from 200 FBM
simulations for each given $L$. The dotted line represents the
expected free space behavior $E_B\sim\Delta$, which is nicely
fulfilled by the data at shorter times and sufficiently large $L$.
At longer times or small $L$ the results show that $E_B$ behaves
very differently when confinement effects are present. The plateau
in $E_B$ is related to the saturation of the curves for the mean
squared displacement, Fig.~\ref{f:msd-fbm}. As the motion is
restricted by the walls roughly above a crossover lag time
$\Delta_{cr}=(L^2/2K_H)^{1/2H}$, the ergodicity breaking parameter
$E_B$ levels off at $\Delta\gtrsim \Delta_{cr}$. The sharp
increase at the end of the curve is due to the singularity when
the lag time reaches the size of the overall measurement time $T$,
which would disappear in the infinite measurement time.

\begin{figure}
\includegraphics[width=0.50\textwidth]{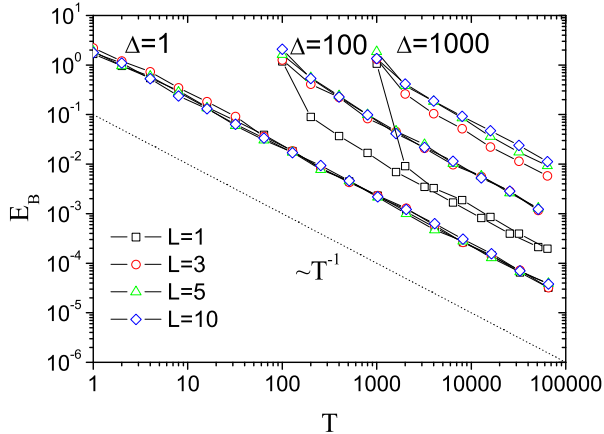}
\caption{(Color online) Ergodicity breaking parameter $E_B$ versus overall measurement time
$T$ at given lag time $\Delta=1$, 100, and 1000. The dotted line depicts a
power law with slope -1, representing the analytic behavior $E_B\simeq T^{
-1}$ in free space. For each $\Delta$, the $E_B$ curves are drawn for different
interval lengths $L=1$, 3, 5, and 10. Each curve was obtained from 200 single
trajectories, and the Hurst exponent was $H=1/4$.}
\label{f:EBt-fbm}
\end{figure}

In Fig.~\ref{f:EBt-fbm} we show $E_B$ for given $\Delta$ as
function of the measurement time $T$ for the same choice of
interval lengths, $L=1,3,5$, and 10. For short lag times $\Delta$,
all $E_B$ curves coincide and decay as $T^{ -1}$, in complete
analogy to the free space motion (dotted line). In the case of
longer $\Delta$ the general trend is that $E_B$ decays like
$T^{-1}$, unaltered with respect to the free case. However, there
is a sudden decrease in $E_B$ for the smallest interval size, for
$L=1$. One can understand this behavior by observing the $E_B$
curve for $L=1$ in Fig.~\ref{f:EB-fbm-lag}; as the fluctuations of
the mean squared displacement are strongly suppressed due to the
tight confinement in this case, $E_B$ has almost no dependence on
$\Delta$ for $\Delta_{cr}\lesssim\Delta \lesssim T$ and the
saturated value is quite small compared to those for other cases.
Therefore, the curves for $L=1$ appear disconnected from the other
curves. Corresponding to the approximate independence of the $L=1$
curve for $\Delta \gtrsim10$ in Fig.~\ref{f:EB-fbm-lag}, we
observe in Fig.~\ref{f:EBt-fbm} that at longer times $T$ the $L=1$
curves approach each other. Only at $T\approx\Delta$ these curves
separate, as then
$\overline{\delta_i^2(T,T)}=[x^H_i(t+T)-x^H_i(t)]^2$, and $E_B$ is
evaluated with the same small number of squared displacement data.
Note that the splitting of the $E_B$ curve can be also observed
for larger $L$ at $\Delta$s larger than $\Delta_{cr}$ under longer
total measurement time $T$ as other $E_B$ curves also have
corresponding constant saturation values for $\Delta
\gtrsim\Delta_{cr}(=(L^2/2K_H)^{1/2H})$ which increases with the
size $L$.

\subsection{Displacement correlation function}

As explained for the stochastic properties of FBM in
Sec.~\ref{sec2}, the position autocorrelation $\langle
x^H(t_1)x^H(t_2)\rangle$ explicitly depends on $t_1$ and $t_2$ as
well as their difference, $|t_1-t_2|$. It is therefore not an
efficient quantity to estimate directly from experimental or
simulations data. However, the correlation function of the
displacements
\begin{equation}
\delta x^H(t,\Delta)=x^H(t+\Delta)-x^H(t)
\end{equation}
depends only on the time interval $\Delta$ of the displacement,
\begin{eqnarray}
\label{eq:jump-fbm}
\langle\delta x^H(t,\Delta)\delta x^H(t-\Delta,\Delta)\rangle=K_H(2^{2H}-2)
\Delta^{2H}
\end{eqnarray}
for free FBM. This relation is derived in App.~\ref{app1}. This quantity is
anticorrelated for $0<H<1/2$ (subdiffusive motion), uncorrelated for $H=1/2$
(normal Brownian motion), and positively correlated for $1/2<H<1$
(superdiffusion). As Eq.~(\ref{eq:jump-fbm}) does not depend on the measurement
time $T$ the value of the ensemble averaged value is identical to the
corresponding time average.

\begin{figure}
\includegraphics[width=0.50\textwidth]{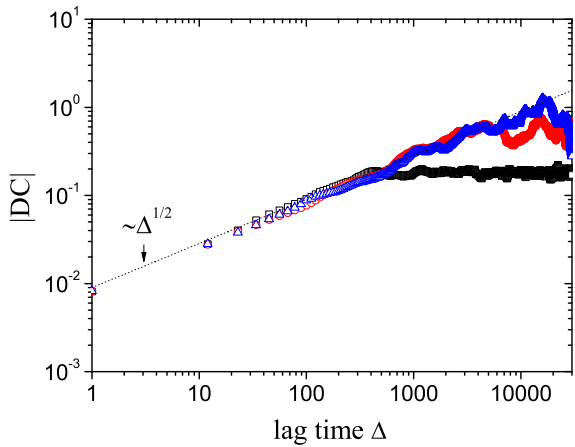}
\caption{(Color online) Absolute value of the displacement correlation (DC) versus
lag time $\Delta$ for $L=1$, 2, and 3 (from bottom to top) with a slope
proportional to $\Delta^{2H}$ (dotted line). Total measurement time is $T=
2^{17}\approx1.3\cdot10^5$, and the Hurst exponent is $H=1/4$.
Each curve was obtained via time averaging from a single particle trajectory. }
\label{f:jump-fbm}
\end{figure}

We present the time averaged displacement correlation for confined
subdiffusive motion ($H=1/4$) in Fig.~\ref{f:jump-fbm}. Because of
the negativity of expression (\ref{eq:jump-fbm}) the absolute
value of the displacement correlation is drawn in the log-log
representation. At short lag times the slope of the correlation
functions is proportional to $2H$ as expected from
Eq.~(\ref{eq:jump-fbm}). However, at long lag times, we
interestingly observe fluctuations of the correlations around a
constant value, reflecting the confinement of the motion.

\subsection{Dimensionality}

\begin{figure}[tb]
\includegraphics[width=0.50\textwidth]{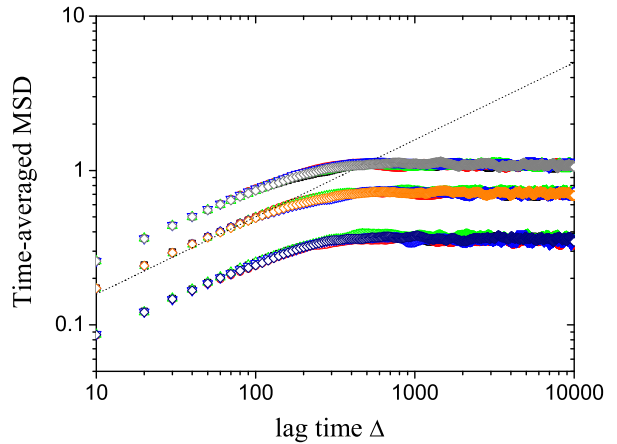}
\caption{(Color online) Time-averaged mean squared displacement (MSD) curves versus lag
time $\Delta$ for $L=1$ in 1D, 2D, and 3D space (from bottom to top), with
a slope $\Delta^{2H}$ (dotted line). For each dimension, 5 trajectories
were drawn with total measurement time $T=2^{17}$ and $H=1/4$.}
\label{f:msd3D-fbm1}
\end{figure}

To mimic the anomalous diffusion of particles inside biological cells, we
also simulate two- and three-dimensional FBM based on Eq.~(\ref{eq:fbm-2d3d})
in the presence of reflecting walls. In free space, the ensemble average of
the time averaged mean squared displacement is simply given by
\begin{eqnarray}
\nonumber
\left<\overline{\delta^2(\Delta,T)}\right>&=&\frac{1}{T-\Delta}\int_0^{T-
\Delta}\left<[\vt{x}^H(t+\Delta)-\vt{x}^H(t)]^2\right>dt,\\
&=&2dK_H\Delta^{2H},
\label{eq:msdfbm3D}
\end{eqnarray}
i.e., it is additive as for the ensemble average. This behavior is indeed
observed in Fig.~\ref{f:msd3D-fbm1} where five different mean squared
displacement curves are drawn for $L=1$ in 1D, 2D, and 3D, respectively. Only
the height of
the curves are affected by the dimensionality. There is no noticeable
difference in the scatter of the curves.

\begin{figure}
\includegraphics[width=0.50\textwidth]{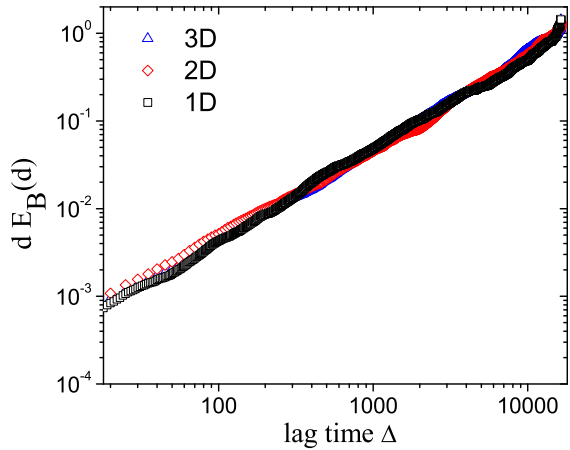}
\caption{(Color online) Rescaled ergodicity breaking parameter $dE_B$ versus lag time $\Delta$ for
interval size $L=5$ for 1D, 2D, and 3D space (from top to bottom), with
Hurst exponent $H=1/4$ and measurement time $T=2^{14}$. $E_B$ was evaluated from 200 trajectories with
different initial positions.}
\label{f:EB3D-fbm2}
\end{figure}

We further investigate the effects of dimensionality on the
scatter of the mean squared displacement curves, as possibly the
strong scatter observed in experiments \cite{golding,lene,elbaum}
may also occur for FBM in higher dimensions. To see the effect of
dimensionality on the ergodicity behavior we measure $E_B$ versus
lag time for one-, two-, and three-dimensional embedding dimension
for the same values of $L$ and $H$. Interestingly, the result
shows that $E_B$ tends to decrease with increasing dimensionality
$d$, meaning that for FBM big scatter is not caused by higher
dimensions in presence of reflecting walls. In fact, from
Eqs.~(\ref{eq:EB}) and (\ref{eq:msdfbm3D}), we can analytically
derive the relation
\begin{equation}
\label{eq:EBD}
E_B(d)=\frac{E_B(d=1)}{d},
\end{equation}
which still holds in the case of confined motion (see appendix B
for the derivation). This relation is numerically confirmed in
Fig.~\ref{f:EB3D-fbm2} where three $E_B$ curves collapse upon
rescaling by $dE_B(d)$. According to this relation, we expect that
ergodic behavior obtained in one-dimensional confined motion
(Figs.~\ref{f:EB-fbm-lag} and \ref{f:EBt-fbm}) will also be
present in multiple dimensions with a factor of $1/d$.

\section{Fractional Langevin equation motion in confined space}
\label{sec5}

In this section we analyze FLE motion under confinement. Due to the different
physical basis compared to FBM, in particular, the occurrence of inertia, we
observe interesting variations on the properties studied in the previous
section.

\subsection{Mean squared displacement}

Using the correlation function \cite{pottier}
\begin{eqnarray}
\nonumber \langle
y(t_1)y(t_2)\rangle&=&\frac{k_BT}{m}[t_1^2E_{2\oH,3}(-\gamma
t_1^{2\oH})+t_2^2E_{2\oH,3}(-\gamma t_2^{2\oH})\nonumber
\\
&&\hspace*{-0.8cm}-(t_2-t_1)^2E_{2\oH,3}(-\gamma
|t_2-t_1|^{2\oH})],
\end{eqnarray}
one can show analytically that, similarly to the FBM, the ensemble
averaged second moment $\langle y^2(t)\rangle$ is identical to its
time averaged analog $\left<\overline{\delta^2(\Delta)}\right>$,
for all $\Delta$ in free space, namely
\begin{equation}
\left<\overline{\delta^2(\Delta,T)}\right>=2\frac{k_BT}{m}\Delta^2E_{2\oH,3}(-\gamma
\Delta^{2\oH}).\label{eq:tmsd-FLE}
\end{equation}
Thus, the time averaged mean squared displacement turns over from a ballistic
motion
\begin{equation}
\left<\overline{\delta^2(\Delta,T)}\right>\simeq\Delta^2
\end{equation}
at short lag time to the subdiffusive behavior
\begin{equation}
\left<\overline{\delta^2(\Delta,T)}\right>\simeq\Delta^{2-2\oH}
\end{equation}
at long lag times, in free space.

\begin{figure}
\includegraphics[width=0.50\textwidth]{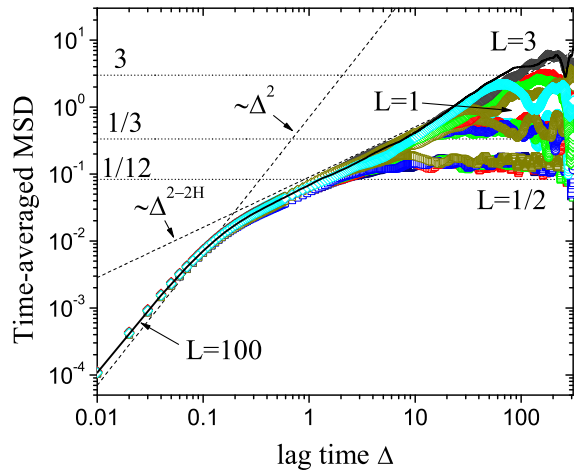}
\caption{(Color online) Time-averaged mean squared displacement (MSD) versus lag time
$\Delta$. The two dashed lines represent the two asymptotic scaling
behaviors $\left<\overline{\delta^2(\Delta,T)}\right>\simeq\Delta^2$ and
$\left<\overline{\delta^2(\Delta,T)}\right>\simeq\Delta^{2-2\oH}$. For each
given $L=1/2$, 1, and 3, five different trajectories are drawn to visualize
the scatter. As a reference curve for motion in free space, the curve for
$L=100$ (line) is drawn. In the
simulation, we chose the Hurst exponent $\oH=5/8$ [NB: for the FLE this means
subdiffusion], time increment $dh=0.01$, particle mass $m=1$, initial velocity
$v_0=1$, initial position $y_0=0$, friction coefficient $\overline{\gamma}=10$,
and $k_BT=1$.}
\label{f:msd-gle}
\end{figure}

We numerically study how this scaling behavior is affected by the
confinement. Figure~\ref{f:msd-gle} shows typical curves for the
time averaged mean squared displacement, for interval sizes
$L=1/2$, 1, 3, and 100 (regarded as free space motion) with
identical initial conditions and Hurst exponent $\oH=5/8$. The
results are summarized as follows: (1) We observe both scaling
behaviors,
$\left<\overline{\delta^2(\Delta,T)}\right>\simeq\Delta^2$ turning
over to $\simeq\Delta^{2-2\oH}$, for confined FLE motions. (2) For
narrow intervals, the curves eventually reach the saturation
plateau within the chosen total measurement time $T$. The
saturation values are approximately $L^2/3$. For interval size
$L=1/2$, the saturated value is noticeably larger than $L^2/3$,
which appears to be caused by multiple reflection events on the
walls. The same behavior is observed in the FBM case when
considering a large value of $H\geq1/2$, or very narrow intervals
for the given $H=1/4$. (3) As in the case of FBM, the scatter
becomes pronounced as the interval length increases.

\subsection{Ergodicity breaking parameter}

\begin{figure}
\includegraphics[width=0.50\textwidth]{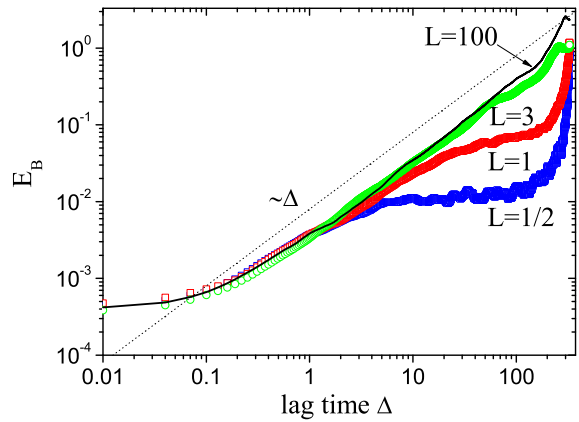}
\caption{(Color online) Ergodicity breaking parameter $E_B$ versus lag time $\Delta$ for
interval sizes $L=1/2$, 1, 3, and 100. The straight line
corresponds to the free space behavior $E_B\simeq\Delta$. Each curve was
obtained from 200 single trajectories, with the same parameter values used
in Fig.~\ref{f:msd-gle}.}
\label{f:EB-gle}
\end{figure}

From a simple argument and simulations it was shown in Ref.~\cite{deng}
that the FLE and FBM mean squared displacements are asymptotically equal,
$\left<\overline{\delta^2(y)}\right>\sim\left<\overline{\delta^2(x^H)}\right>$,
similarly for the ergodicity breaking parameter, $E_B(y)\sim E_B(x^H)$. Here
the asymptotic equivalence is valid at long measurement times $T$, and the
derivation holds for motion in free space. From 200 trajectories of the mean
squared displacement we measure the ergodicity breaking parameter $E_B$ as
function of lag time $\Delta$ for
interval lengths $L=1/2$, 1, 3, and 100 in Fig.~\ref{f:EB-gle}. The behavior
is similar to the corresponding curves for FBM, displayed in
Fig.~\ref{f:EB-fbm-lag}: $E_B$ significantly deviates from the reference curve
for free space motion (i.e., the longer $\Delta$ behavior for $L=100$ and the
drawn power law $\simeq\Delta$), due to the confinement effect. $E_B$ tends to
decrease with smaller $L$ for the same value of $\Delta$. However, the plateau
at short lag times that is still observed for $L=100$ (regarded as free space
motion), is due to the initial ballistic motion of FLE. In that regime the
initially directed motion renders the random noise effect negligible.

\subsection{Displacement correlation function}

Using the correlation function $\langle y(t_1)y(t_2)\rangle$, we
analytically obtain the displacement correlation function in free
space in the form (refer to App.~\ref{app1} for the derivation)
\begin{eqnarray}
\nonumber
\langle\delta y(t,\Delta)\delta y(t-\Delta,\Delta)\rangle&=&4\frac{k_BT}{m}\Delta^2E_{
2\oH,3}[-\gamma(2\Delta)^{2\oH}]\\
&&\hspace*{-0.8cm}-2\frac{k_BT}{m}\Delta^2E_{2\oH,3}[-\gamma\Delta^{2\oH}],
\label{eq:jump_gle}
\end{eqnarray}
so that we observe the following asymptotic behavior
\begin{eqnarray}
\nonumber
\langle\delta y(t,\Delta)\delta y(t-\Delta,\Delta)\rangle&&\\[0.2cm]
&&\hspace*{-2.8cm}\sim\left\{
\begin{array}{ll}\displaystyle\frac{2k_BT}{m\Gamma(3)}\Delta^2 & \mbox{for }
\Delta\rightarrow 0,\\[0.4cm]
\displaystyle\frac{(2^{2-2H}-2)k_BT}{m\gamma \Gamma(3-2\oH)}\Delta^{2-2\oH} &
\mbox{for }\Delta\rightarrow\infty,
\end{array}\right.
\label{eq:jump_gle2}
\end{eqnarray}
where $\delta y(t,\Delta)=y(t+\Delta)-y(t)$. Above expression
shows that the displacement correlation has two distinct scaling
behaviors. At short lag times, it grows like $\Delta^2$ and is
positive, due to the ballistic motion. At long lag times, it is
negative in the domain $1/2<\oH<1$, exhibiting the same
subdiffusive behavior as observed for FBM
[cf.~Eq.~(\ref{eq:jump-fbm})] when we replace
$H\rightarrow2-2\oH$. Note that to bridge these two scaling
behaviors the displacement correlation passes the zero axis at
$\Delta= \Delta_c$ that satisfies
$2E_{2\oH,3}\left[-\gamma(2\Delta_c)^{2\oH}\right]=E_{2\oH,
3}\left[-\gamma\Delta_c^{2\oH}\right]$ in free space. For small
$\gamma$, we find approximately
\begin{equation}
\Delta_c\approx\left(\frac{\Gamma(2\oH+3)}{2\Gamma(2\oH-1)(2^{2\oH+1}-1)}\frac{m}{
\overline{\gamma}}\right)^{1/2\oH},
\end{equation}
such that it becomes exactly the momentum relaxation time
$m/\overline{ \gamma}$ for normal Brownian motion ($\oH=1/2$). In
the limit $\oH\rightarrow1$, $\Delta_c$ goes to infinity to
satisfy the equality $2E_{2\oH,3}\left[-\gamma(
2\Delta_c)^{2\oH}\right]=E_{2\oH,3}\left[-\gamma\Delta_c^{2\oH}\right]$.
Thus, $\Delta_c$ can be interpreted as the typical timescale for
the persistence of the ballistic motion.

\begin{figure}
\includegraphics[width=0.50\textwidth]{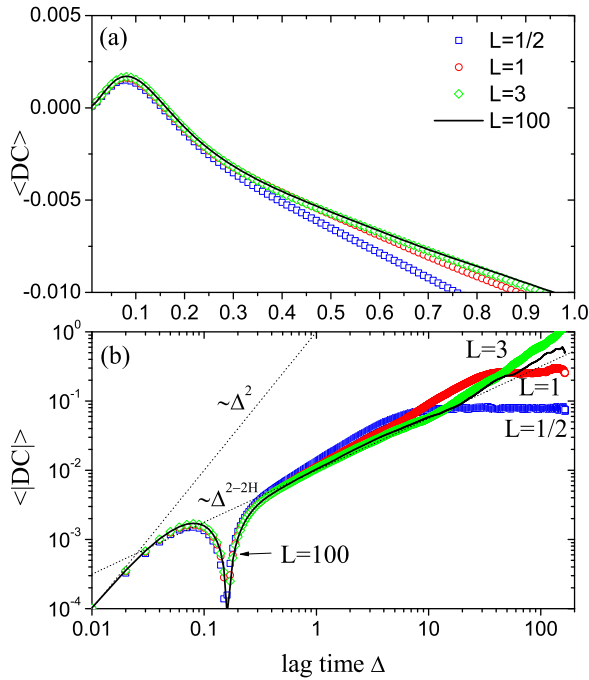}
\caption{(Color online) (a) Displacement correlation (DC) versus lag time $\Delta$
for $L=1/2$, 1, 3, and 100 (from bottom to top). (b) Absolute value of the
displacement correlation as a function of $\Delta$ for the given values
of $L$. The two slopes correspond to the limiting behavior $\Delta^{2}$ and
$\Delta^{2-2\oH}$. In (a) and (b), the ensemble averaged curves were obtained
from 200 different time averaged displacement correlation curves. Same parameter
values as in Fig.~\ref{f:msd-gle}.}
\label{f:jump-gle}
\end{figure}

Figure~\ref{f:jump-gle} shows (a) the displacement correlation
versus lag time $\Delta$, and (b) the absolute value of the
displacement correlation as function of $\Delta$, for $L=1/2$, 1,
3, and 100. The scaling properties derived in
Eq.~(\ref{eq:jump_gle2}) are indeed observed. At short lag times
all curves are positive and scale like $\sim\Delta^2$, before
decreasing to zero. In the long lag time regime the displacement
correlation becomes negative and the predicted scaling behavior
$\simeq \Delta^{2-2\oH}$ is observed. For small intervals ($L=1/2$
and 1), it is saturated due to the confinement effect as seen in
the case of FBM.

\subsection{Dimensionality}

In the case when the memory tensor is diagonalized, each coordinate motion is
independent and FLE motion exhibits qualitatively the same behavior as shown
in the case of FBM with increasing dimensionality. From the mean squared
displacement curves, the same scaling behavior is expected with more
elevated amplitude for
higher dimensionality. In fact, when each coordinate motion is decoupled, a
$d$-dimensional motion effectively increases the number of
single trajectories $d$ times compared to the one-dimensional case. Therefore,
the scatter in the mean square displacement curves decreases with increasing
dimensionality and the ergodicity breaking parameter $E_B$ is expected to
follow the relation Eq.~(\ref{eq:EBD}).

\section{Conclusion}
\label{sec6}

Motivated by recent single particle tracking experiments in biological cells,
in which confinement due to the rather small cell size becomes relevant, we
studied FBM and FLE motions in confined space. In particular we analyzed the
effects of confinement and dimensionality on the stochastic and ergodic
properties of the two processes. Interestingly for both stochastic models,
the confinement tends to decrease the value of the ergodicity breaking
parameter $E_B$ compared to that in free space. The same trend is observed
for increasing dimensionality. Correspondingly the scatter of time averaged
quantities between individual trajectories is quite small, apart from regimes
when the lag time $\Delta$ becomes close to the overall measurement time $T$
and the sampling statistics for the corresponding time average become poor.
The relaxation of the ergodicity breaking parameter as function of measurement
time is quite similar to previous results in free space. We conclude that
neither confinement nor dimensionality effects lead to the appearance of
significant ergodicity breaking or scatter between single trajectories.

The displacement correlation function introduced here is a useful
quantity that can be easily obtained from single particle
trajectories. It can be used as a tool to discriminate one
stochastic model from another. For subdiffusive motion governed by
FBM and FLE motion, the displacement correlation should be
negative and saturate in the long measurement time limit due to
the confinement. Notably, the negative decrease
($\sim-\Delta^{\alpha}$) with lag time $\Delta$ and anomalous
diffusion exponent $\alpha$ is an intrinsic property of FBM and
FLE displacement correlations which is clearly distinguished from
that of CTRW subdiffusion. In the latter case, the subdiffusive
motion occurs due to the long waiting time distribution between
successive jumps and there is no spatial correlation between them,
so that displacement correlation only fluctuates around zero
with time. FLE motion can be distinguished from FBM motion since
the displacement correlation has a positive value at short times
due to the ballistic motion in the FLE model.

\acknowledgments

We thank Stas Burov and Eli Barkai for helpful and enjoyable discussions.
Financial support from the DFG is acknowledged.

\appendix

\section{Derivation of the displacement correlation function}
\label{app1}

In this appendix we derive analytical expressions for the
displacement correlations, Eqs.~(\ref{eq:jump-fbm}) and
(\ref{eq:jump_gle}). For a stochastic variable $x$, we define
\begin{equation}
\delta x(t,\Delta)=x(t+\Delta)-x(t).
\end{equation}
The displacement correlation is then given by
\begin{eqnarray}
\nonumber
&&\langle\delta x(t,\Delta)\delta x(t-\Delta,\Delta)\rangle=\\
\nonumber
&&\hspace*{0.8cm}\langle x(t+2\Delta)x(t+\Delta)\rangle-\langle x(t+2\Delta)
x(t)\rangle\\
&&\hspace*{0.8cm}-\langle x(t+\Delta)^2\rangle+\langle x(t+\Delta)x(t)\rangle.
\end{eqnarray}
We now calculate this expression for FBM and FLE motions.

\subsection{FBM}

For FBM ($x(t)=x^H(t)$), we use the expression
\begin{equation}
\langle x(t_1)x(t_2)\rangle=K_H(t_1^{2H}+t_2^{2H}-|t_2-t_1|^{2H})
\end{equation}
for the autocorrelation. With this we readily obtain the result
\begin{eqnarray}
\langle\delta x(t,\Delta)\delta x(t-\Delta,\Delta)\rangle=K_H(2^{2H}-2)
\Delta^{2H}.
\end{eqnarray}

\subsection{FLE}

For FLE ($x(t)=y(t)$), we use the correlation function \cite{pottier}
\begin{eqnarray}
\nonumber \langle x(t_1)x(t_2)\rangle&=&
\hspace*{-0.1cm}\frac{k_BT}{m}[t_1^2E_{2\oH,3}(-\gamma
t_1^{2\oH})+t_2^2E_{2\oH,3}(-\gamma t_2^{2\oH})\nonumber \\
&&\hspace*{-0.8cm}-(t_2-t_1)^2E_{2\oH,3}(-\gamma
|t_2-t_1|^{2\oH})].
\end{eqnarray}
The displacement correlation is then obtained as
\begin{eqnarray}
\nonumber
&&\langle\delta x(t,\Delta)\delta x(t-\Delta,\Delta)\rangle\\
\nonumber
&&\hspace*{1.2cm}=\frac{4k_BT}{m}\Delta^2E_{2\oH,3}\left[-\gamma(2\Delta)^{2\oH}\right]\\
&&\hspace*{1.2cm}-\frac{2k_BT}{m}\Delta^2E_{2\oH,3}[-\gamma\Delta^{2\oH}].
\end{eqnarray}
Expanding the generalized Mittag-Leffler function
$E_{2\oH,3}(x)\approx1/ \Gamma(3)+x/\Gamma(2\oH+3)+\cdots$ for
$x\ll1$ the displacement correlation is approximated as
\begin{equation}
\langle\delta x(t,\Delta)\delta x(t-\Delta,\Delta)\rangle\sim\frac{2}{
\Gamma(3)}\frac{k_BT}{m}\Delta^2
\end{equation}
at short lag times. With the expansion
$E_{2\oH,3}(-x)\approx1/x\Gamma(3-2\oH)$ for $x\gg1$ the long lag
time behavior of the displacement correlation is obtained as
\begin{equation}
\langle\delta x(t,\Delta)\delta x(t-\Delta,\Delta)\rangle\sim\frac{2^{2-2\oH}-2}{
\gamma\Gamma(3-2\oH)}\frac{k_BT}{m}\Delta^{2-2\oH}.
\end{equation}
Note that the prefactor $(2^{2-2\oH}-2)/\Gamma(3-2\oH)$ is zero
for $\oH=1/2$ and then becomes increasingly negative, saturating
at the value 1 for $\oH=1$.

\section{Derivation of Equation (\ref{eq:EBD})}

From the definition of the time averaged mean squared displacement,
Eq.~(\ref{eq:msdfbm3D}), we expand $\left<\left(\overline{\delta^2(\Delta,T)}
\right)^2\right>$ in the form
\begin{widetext}
\begin{eqnarray}
\nonumber
\left<\left(\overline{\delta^2(\Delta,T)}\right)^2\right>=\frac{1}{(T-\Delta)
^2}\int_0^{T-\Delta}\int_0^{T-\Delta}&\Big\{&\sum_{i=1}^d\langle[x_i
^H(t_1+\Delta)-x_i^H(t_1)]^2[x_i^H(t_2+\Delta)-x_i^H(t_2)]^2\rangle\\
&&\hspace*{-1.2cm}
+\sum_{i\neq j}\langle[x_i^H(t_1+\Delta)-x_i^H(t_1)]^2\rangle
\langle[x_j^H(t_2+\Delta)-x_j^H(t_2)]^2\rangle\Big\}dt_1dt_2.
\label{eq:del2}
\end{eqnarray}
Using the Isserlis theorem for Gaussian process with zero mean \cite{coffey}:
\begin{equation}
\langle x(t_1)x(t_2)x(t_3)x(t_4)\rangle=\langle x(t_1)x(t_2)\rangle\langle
x(t_3)x(t_4) \rangle+\langle x(t_1)x(t_3)\rangle\langle x(t_2)x(t_4)\rangle
+\langle x(t_1)x(t_4)\rangle\langle x(t_2)x(t_3)\rangle,
\end{equation}
the first term in the braces in Eq.~(\ref{eq:del2}) can be rewritten as
\begin{eqnarray}
\nonumber
&&\sum_{i=1}^d\langle[x_i^H(t_1+\Delta)-x_i^H(t_1)]^2 [x_i^H(t_2+\Delta)-
x_i^H(t_2)]^2\rangle
=\sum_{i=1}^d\langle[x_i^H(t_1+\Delta)-x_i^H(t_1)]^2 \rangle\langle[x_i^H
(t_2+\Delta)-x_i^H(t_2)]^2\rangle\\
&&\hspace*{6.8cm}
+2\sum_{i=1}^d\langle[x_i^H(t_1+\Delta)-x_i^H(t_1)][x_i^H(t_2+\Delta)-x_i^H
(t_2)]\rangle^2.
\label{eq:expand}
\end{eqnarray}
In this expression, we note that the sum of the second term in
Eq.~(\ref{eq:del2}) and the first term in Eq.~(\ref{eq:expand}) yields
$\left<\overline{\delta^2(\Delta,T)}\right>^2$:
\begin{eqnarray}
\nonumber
\left<\overline{\delta^2(\Delta,T)}\right>^2&=&\frac{d^2}{(T-\Delta)^2}\int_0^
{T-\Delta}\int_0^{T-\Delta}\langle[x^H(t_1+\Delta)-x^H(t_1)]^2\rangle
\langle[x^H(t_2+\Delta)-x^H(t_2)]^2\rangle dt_1dt_2\\[0.4cm]
&=&d^2\langle\overline{\delta^2}\rangle_{1D}^2,
\label{eq:delsquare}
\end{eqnarray}
where we used the property
\begin{equation}
\sum_{i,j}\langle[x_i^H(t_1+\Delta)-x_i^H(t_1)]^2\rangle\langle[x_j^H(t_2+
\Delta)-x_j^H(t_2)]^2\rangle =d^2\langle[x^H(t_1+\Delta)-x^H(t_1)]^2\rangle
\langle[x^H(t_2+\Delta)-x^H(t_2)]^2\rangle
\end{equation}
due to the independence of the motion in each coordinate direction. We also
note that the expression $\left<\left(\overline{\delta^2(\Delta,T)}\right)^2
\right>-\left<\overline{\delta^2(\Delta,T)}\right>^2$ simplifies to
\begin{eqnarray}
\nonumber
\left<(\overline{\delta^2(\Delta,T})^2\right>-\langle\overline{\delta^2}
\rangle^2&=&\frac{2}{(T-\Delta)^2}\sum_{i=1}^{d}\int_0^{T-\Delta}\int_0^{T-
\Delta}\langle[x_i^H(t_1+\Delta)-x_i^H(t_1)]\times[x_i^H(t_2+\Delta)-x_i^H
(t_2)]\rangle^2dt_1dt_2\\
&=&d[\langle(\overline{\delta^2})^2\rangle_{1D}-\langle\overline{\delta^2}
\rangle_{1D}^2].
\label{eq:delvariance}
\end{eqnarray}
From Eqs.~(\ref{eq:delsquare}) and (\ref{eq:delvariance}), the ergodicity
breaking parameter follows the general relation
\begin{equation}
E_B(d)=\frac{\left<\left(\overline{\delta^2(\Delta,T)}\right)^2\right>-
\left<\overline{\delta^2(\Delta,T)}\right>^2}{\left<\overline{\delta^2(
\Delta,T)}\right>^2}=\frac{E_B(d=1)}{d}.
\end{equation}
\end{widetext}


\begin{thebibliography}{99}

\bibitem{report} R. Metzler and J. Klafter, Phys. Rep. {\bf 339} 1 (2000);
J. Phys. A \textbf{37}, R161 (2004).

\bibitem{scher} H. Scher and E. W. Montroll, Phys. Rev. B {\bf 12}, 2455 (1975).

\bibitem{grl} J. W. Kirchner, X. Feng, and C. Neal, Nature \textbf{403}, 524
(2000); H. Scher, G. Margolin, R. Metzler, J. Klafter, and B. Berkowitz,
Geophys. Res. Lett. \textbf{29}, 1061 (2002).

\bibitem{mainardi} F. Mainardi, M. Raberto, R. Gorenflo, and E. Scalas, Physica
A \textbf{287}, 468 (2000).

\bibitem{weitz} I. Y. Wong, M. L. Gardel, D. R. Reichman, E. R. Weeks,
M. T. Valentine, A. R. Bausch, and D. A. Weitz, Phys. Rev. Lett. \textbf{92},
178101 (2004).

\bibitem{kimmich} E. Fischer, R. Kimmich, and N. Fatkullin,
J. Chem. Phys. \textbf{104}, 9174 (1996).

\bibitem{weeks} E. R. Weeks, J. C. Crocker, A. C. Levitt, A. Schofield, and
D. A. Weitz, Science \textbf{287}, 627 (2000).

\bibitem{golding} I. Golding and E. C. Cox, Phys. Rev. Lett. {\bf 96}, 098102
(2006).

\bibitem{lene} I. M. Toli{\'c}-N{\o}rrelykke, E. L. Munteanu, G. Thon, L.
Oddershede, and K. Berg-S{\o}rensen, Phys. Rev. Lett. {\bf 93}, 078102 (2004).

\bibitem{elbaum} A. Caspi, R. Granek, and M. Elbaum, Phys. Rev. Lett.
\textbf{85}, 5655 (2000); Phys. Rev. E. {\bf 66}, 011916 (2002).

\bibitem{seisenhuber} G. Seisenberger, M. U. Ried, T. Endre{\ss}, H.
B{\"u}ning, M. Hallek, and C. Br{\"a}uchle, Science \textbf{294}, 1929
(2001).

\bibitem{garini} I. Bronstein, Y. Israel, E. Kepten, S. Mai, Y. Shav-Tal,
E. Barkai, and Y. Garini, Phys. Rev. Lett. \textbf{103}, 018102 (2009).

\bibitem{weiss} M. Weiss, M. Elsner, F. Kartberg, and T. Nilsson, Biophys. J.
\textbf{87}, 3518 (2004); M. Weiss, H. Hashimoto, and T. Nilsson, \emph{ibid.}
\textbf{84}, 4043 (2003). Note: HeLa cells belong to an immortal cell line derived from cancer cells originally taken from Henrietta Lacks in 1951.

\bibitem{vincent} V. Tejedor and R. Metzler (unpublished).

\bibitem{havlin} S. Havlin and D. ben-Avraham, Adv. Physics \textbf{36},
695 (1987).

\bibitem{kimmich1} A. Klemm, R. Metzler, and R. Kimmich, Phys. Rev. E
\textbf{65}, 021112 (2002).

\bibitem{molchan} G. M. Molchan, Commun. Math. Phys. \textbf{205}, 97 (1999).

\bibitem{goychuck}I. Goychuk and P. H\"{a}nggi, Phys. Rev. Lett.
{\bf 99}, 200601 (2007).

\bibitem{chaudhury}S. Chaudhury and B. J. Cherayil, J. Chem.
Phys. {\bf 125}, 024906 (2006); S. Chaudhury, D. Chatterjee, and
B. J. Cherayil, \emph{ibid}. {\bf 129}, 075104 (2008).

\bibitem{chechkin} O. Yu. Slyusarenko, V. Yu. Gonchar, A. V. Chechkin,
I. M. Sokolov, and R. Metzler (unpublished).

\bibitem{goychuck2} I. Goychuk, Phys. Rev. E {\bf 80}, 046125 (2009).

\bibitem{stas1} S. Burov and E. Barkai, Phys. Rev. Lett. \textbf{100},
070601 (2008).

\bibitem{YHe} Y. He, S. Burov, R. Metzler, and E. Barkai, Phys. Rev. Lett.
{\bf 101}, 058101 (2008); A. Lubelski, I. M. Sokolov, and J. Klafter, \emph{ibid.} \textbf{100}, 250602 (2008).

\bibitem{app} R. Metzler, V. Tejedor, J.-H. Jeon, Y. He, W. Deng, S. Burov,
and E. Barkai, Acta Phys. Polon. B \textbf{40}, 1315 (2009).

\bibitem{deng} W. Deng and E. Barkai, Phys. Rev. E {\bf 79}, 011112 (2009).

\bibitem{bao} J.-D. Bao, P. H\"anggi, and Y.-Z. Zhuo, Phys. Rev. E. {\bf 72},
061107 (2005).

\bibitem{stas} S. Burov, R. Metzler, and E. Barkai (unpublished).

\bibitem{thomas} T. Neusius, I. M. Sokolov, and J. C. Smith, Phy. Rev. E
{\bf 80}, 011109 (2009).

\bibitem{kolmogorov} A. N. Kolmogorov, Doklady Akademii Nauk SSSR (N.S.)
\textbf{26}, 115 (1940).

\bibitem{yaglom} A. M. Yaglom, American Mathematical Society Translations
Series 2 \textbf{8}, 87 (1958).

\bibitem{mandelbrot1} B. B. Mandelbrot, Comptes Rendus (Paris) \textbf{260},
3274 (1965).

\bibitem{mandelbrot} B. B. Mandelbrot and J. W. van Ness, SIAM Rev. \textbf{10},
422 (1968).

\bibitem{hurst} H. E. Hurst, Trans. Am. Soc. Civ. Eng. \textbf{116}, 770
(1951); H. E. Hurst, R. O. Black, and Y. M. Simaika, Long term storage: an
experimental study (Constable, London, 1965).

\bibitem{finance} I. Adelman, Amer. Econom. Rev. \textbf{60}, 444 (1965);
C. W. J. Granger, Econometrica \textbf{34}, 150 (1966).

\bibitem{weiss1} J. Szymanski and M. Weiss, Phys. Rev. Lett. \textbf{103},
038102 (2009).

\bibitem{vincent1} V. Tejedor, O. B{\'e}nichou, R. Voituriez, R. Jungmann,
F. Simmel, C. Selhuber, L. Oddershede, and R. Metzler (to appear in Biophs. J.).

\bibitem{oksendal} F. Biagini, Y. Hu, B. {\O}ksendal, and T. Zhang,
Stochastic calculus for fractional Brownian motion and applications
(Springer, Berlin, 2008).

\bibitem{weron} A. Weron and M. Magdziarz, Euro. Phy. Lett. \textbf{86}, 60010 (2009).

\bibitem{qian2} H. Qian, \emph{Process with Long-Range
Correlations: Theory and Applications}, Lecture Notes in Physics
Vol. 621, edited by G. Rangarajan and M. Z. Ding (Springer, New
York, 2003).

\bibitem{xie} S. C. Kou and X. S. Xie, Phys. Rev. Lett. \textbf{93}, 180603
(2004).

\bibitem{FGNcomment} The autocorrelation function for
$0<H<1/2$ becomes positive at $t_1=t_2$ due to the second term in
Eq.~(\ref{xi}) and has the property $\int_{-\infty}^{\infty}dt
\langle \xi^H(t)\xi^H(0)\rangle=0$.

\bibitem{feder} J. Feder, Fractals (Plenum Press, New York, 1988);
B. B. Mandelbrot, The fractal geometry of nature (W. H. Freeman and Company,
New York, 1977).

\bibitem{unterberger} J. Unterberger, Ann. Prob. \textbf{37}, 565 (2009).

\bibitem{qian} H. Qian,  G. M. Raymond, and J. B. Bassingthwaighte, J. Phys.
A \textbf{31}, L527 (1998).

\bibitem{falconer} K. Falconer, Fractal geometry: mathematical
foundations and applications (Wiley, Chichester, UK, 1990).

\bibitem{langevin} P. Langevin, Comptes Rendus {\bf 146}, 530 (1908).

\bibitem{vankampen} N. G. van Kampen, Stochastic Processes
in Physics and Chemistry (North-Holland, Amsterdam, 1981).

\bibitem{coffey} W. T. Coffey, Y. P. Kalmykov, and J. T. Waldron, The Langevin
equation: with applications to stochastic problems in physics, chemistry and
electrical engineering, second edition (World Scientific, Singapore, 2003).

\bibitem{zwanzig} R. Zwanzig, Nonequilibrium Statistical Mechanics
(Oxford University Press, Oxford, UK, 2001).

\bibitem{kubo} R. Kubo in Tokyo lectures in theoretical physics, edited by
R. Kubo (W. A. Benjamin, Inc., New York, NY, 1966).

\bibitem{berne} B. J. Berne, J. P. Boon, and S. A. Rice, J. Chem. Phys.
\textbf{45}, 1086 (1966).

\bibitem{podlubny} I. Podlubny, Fractional differential equations (Academic
Press, New York, 1998).

\bibitem{bateman} A. Erd{\'e}lyi, editor, Bateman Manuscript Project: Higher
Transcendental Functions, Vol. III (McGraw-Hill Book Co., New
York, NY, 1955).

\bibitem{lutz} E. Lutz, Phys. Rev. E. {\bf 64}, 051106 (2001).

\bibitem{pottier} N. Pottier, Physica A {\bf 317}, 371 (2003).

%
%
%
%




\bibitem{hosking} J. R. M. Hosking, Water Res. Res. {\bf 20} 1898 (1984).

\bibitem{kai} K. Diethelm, N. J. Ford, and A. D. Freed, Nonlinear Dynamics
\textbf{29}, 3 (2002).



\end{thebibliography}
\end{document}